\documentclass[11pt]{article}
\usepackage{epsfig,amsmath,amsfonts,amssymb,amsthm}
\DeclareFixedFont{\eightrm}{OT1}{cmr}{m}{n}{10pt}

\def\a{\alpha}

\def\Ga{\Gamma}

\def\la{\lambda}

\def\m{\mu}
\def\n{\nu}

\def\th{\theta}
\def\eps{\epsilon}
\def\ee{\varepsilon}

\newcommand{\vf}{\varphi}
\newcommand{\R}{\mathbb{R}}
\newcommand{\del}{\partial}

\newcommand{\thalf}{\tfrac{1}{2}}             
\def\Tr{{\rm Tr}}
\newcommand{\tri}{\Delta}                     
\newcommand{\sepword}[1]{\quad\mbox{#1}\quad} 

\def\ds{\displaystyle}

\begin{document}
\thispagestyle{empty}
\begin{titlepage}
\rightline{CPT-2004/P.136}
\rightline{UCM-FTI-04/121}

\vglue 60pt
\centerline{\Large \bf Trouble with space-like
               noncommutative field theory}
\vskip 40pt
\centerline {\Large V. Gayral${}^{a,}$\footnote{Also at Universit\'e de
              Provence, gayral@cpt.univ-mrs.fr}, 
              J.~M. Gracia-Bond\'{\i}a${}^b$ and F. Ruiz Ruiz${}^b$}

\vskip 10pt
\begin{center}
{\it $\!{}^a$ Centre de Physique Th\'eorique, UMR 6207, 13288
     Marseille, France.\\
${}^b$ Departamento de F\'{\i}sica Te\'orica I,
       Universidad Complutense de Madrid, 28040 Madrid, Spain.}
\end{center}
\vskip 100pt

{\leftskip=40pt \rightskip=40pt
It is argued that the one-loop effective action for space-like
noncommutative (i) $\lambda\varphi^4$ scalar field theory and (ii) 
$U(1)$ gauge theory
does not exist. This indicates that such theories are not
renormalizable already at one-loop order and suggests
supersymmetrization and reinvestigating other types of
noncommutativity.

\par }

\vfill
\rightline{\it Dedicated to A. Galindo on his 70th birthday}
\vskip 40pt
\end{titlepage}
\newpage
\setcounter{page}{2}


\vskip 40pt
The consistency of a noncommutative quantum field model crucially
depends, even at one loop, on the choice of the matrix
$\Theta:=(\Theta^{\m\n})$ of noncommutative parameters. As is
well-known, in a noncommutative field theory with superficial UV
divergences, one-loop radiative corrections to 1PI Green functions
consist of a planar part and a nonplanar part. The planar part
contains the local UV divergences and has the same dependence on
$\Theta$ as the classical action, whereas the nonplanar part is UV
finite, but depends on $\Theta$ in a complicated way. By adding
suitable local counterterms to the classical action, the UV
divergences in the planar parts are subtracted and one is thus left
with renormalized Green functions.

These functions must be consistent with unitarity. However, combining
renormalizability and unitarity is not a trivial undertaking. It is
known that for space-like or `magnetic' $\Theta$ the one-loop
renormalized Green functions preserve unitarity \cite{Gomis}. Indeed,
since the time-space components $\Theta^{0i}$ of a space-like $\Theta$
vanish, no violations of unitarity are introduced when using Wick
rotation to compute Feynman integrals. The problem that arises then is
that the renormalized 1PI Green functions, while being consistent with
unitarity, in general do not yield a well-defined effective action,
thus putting in jeopardy the existence of space-like noncommutative
quantum field theories. In this paper we show that this is the case
for a real scalar field theory and $U(1)$ gauge theory, and discuss
the connections with other issues in the literature.

We derive this result in configuration space, in which it comes out in
a very elegant and natural way, and then translate it into terms of
the known expressions for the 1PI Green functions in momentum space.
It is important to mention at the outset that we are able to recover
for the latter functions the results already in the literature. The
point is that they do not define an effective action.

Let us consider a real field $\vf(x)$ with classical action
\begin{equation*}
         S[\vf] = \int d^4\!x\, \biggl(\, - \frac{1}{2}\vf\,\Box\,\vf\
         + \frac{1}{2}\,m^2\vf^2
         + \frac{\lambda}{4!} \,\vf\star\vf\star\vf\star\vf\bigg)~,
\end{equation*}
where $\Box=\del^\m\del_\m$ is the d'Alambertian, $m^2$ is the mass
squared, $\lambda$ is the coupling constant and $\star$ denotes the
Moyal product. We consider space-like noncommutativity, for which the
components $\Theta^{0i}$ vanish $(i=1,2,3)$. Without loss of
generality, by a rotation, $\Theta$ can be put in the canonical
form
\begin{equation*}
     \Theta=\begin{pmatrix} 0 & 0 \\
                            0 & \theta S
            \end{pmatrix}, ~~~{\rm with}~~~
      S = \begin{pmatrix} 0 & 1 \\
                          -1 & 0
           \end{pmatrix}~.
\end{equation*}
It is then convenient to split the coordinates $x$ of a point in
spacetime as $x=(\tilde{x},\bar{x})$, with $\tilde{x}=(x^0,x^1)$ and
$\bar{x}=(x^2,x^3)$. Similarly, we write $p=(\tilde p,\bar p)$ in
momentum space. Here the Moyal product of two functions $f$ and $g$
is given by
\begin{equation}
     f\star g\,(x) = \frac{1}{(2\pi)^2}
        \int d^2\!\bar{u}\,\, d^2\!\bar{v} \, \,
          {\rm e}^{-i\bar{u}\cdot\bar{v}}
            f\big(\tilde{x},\bar{x}-\thalf \theta S\bar{u}\big)\,\,
            g(\tilde{x},\bar{x}+\bar{v})~.
\label{Rieffeldef}
\end{equation}
This definition, a particular case of one due to
Rieffel~\cite{Rieffel}, is equivalent to the more familiar one
\begin{displaymath}
f\star g\,(x) = f(x)\, \exp\Big( \frac{i}{2}~\theta\,
       \ee^{\m\n}\, \overleftarrow {\partial_{\bar{x}^\m}} ~
         \overrightarrow{\partial_{\bar{x}^\n}}\Big) \, g(x)~,
\end{displaymath}
under conditions spelled in~\cite{Nereid}.

Using standard techniques and after Wick rotating the time coordinate,
the one-loop contribution $\Gamma_1$ to the effective action can be
recast~\cite{Zinn-Justin} as
\begin{equation*}
     \Gamma_1[\vf] = -\,\frac{1}{2} \int_0^\infty \frac{dt}{t}\, \Tr\,
       \big(\, {\rm e}^{-tH} - {\rm e}^{-tH_0}\, \big),
\end{equation*}
where the operators $H$ and $H_0$ are given by
\begin{equation*}
      H = H_0 + V \sepword{with}
      H_0 = \tri + m^2 \sepword{and}
      V= \frac{\la}{6}\,\,\bigl( L_{\vf\star\vf} + R_{\vf\star\vf}
                         + L_{\vf}\,R_{\vf}\bigr)\,.
\end{equation*}
Here $\tri=-\del^\mu\del_\mu$ is the Laplacian in four-dimensional
euclidean space and $L_{f}$ and $R_{f}$ denote the operators of left
and right Moyal multiplication by $f$. Note that, as we have defined
it, the Laplacian $\tri$ is positive definite, also that $L_f$ and
$R_f$ commute with each other, and that $V$ becomes in the commutative
limit the ordinary multiplication operator by $\la\vf^2/2$. The
expression for $\Gamma_1$ above needs regularization. We regularize it
by using a zeta regularization-like method which amounts to replacing
$1/t$ in the integrand with $\mu^{2\eps}/t^{1-\eps}$, where $\eps$ is
complex and $\mu$ is a mass scale introduced to keep the correct mass
dimension. Thus we write for the regularized one-loop contribution to
the effective action
\begin{equation*}
     \Gamma_1^{\rm reg}[\vf] = -\frac{\mu^{2\eps}}{2}\, \int_0^\infty
       \frac{dt}{t^{1-\eps}} ~
         \Tr\, \big({\rm e}^{-tH} - {\rm e}^{-tH_0} \big) \,.
\end{equation*}
Our concern here is to calculate the divergent part of $\Gamma^{\rm
reg}_1[\vf]$ as $\,\eps\to0$, to see whether the resulting
divergences can be subtracted by local counterterms, and if the result
of doing so is well-defined. To calculate the divergent parts as
$\eps\to0$, we proceed in two steps. First, we use the
covariant perturbation method of Barvinsky and
Vilkovisky~\cite{covariant} to compute
\begin{equation}
     \Gamma_1^{\rm reg}[\vf] = -\frac{\,\mu^{2\eps}}{2}\, \int_0^\infty
       \frac{dt}{t^{1-\eps}} ~\sum_{n=1}^\infty ~ K_n(\tri,V,t)~,
\label{expansion}
\end{equation}
where $K_n(\tri,V,t)$ is given by
\begin{equation*}
      K_n(\tri,V,t) = \frac{(-t)^n}{n} \int_0^1 d\a_1\cdots d\a_n\,\,
        \delta\Big(1 - \sum_1^n\a_i\Big)\,\Tr\,
        \bigl[V\,{\rm e}^{-t\a_{1\!} H_0}
	 \,V\,{\rm e}^{-t\a_{2\!} H_0}\,\cdots
           \,V\, {\rm e}^{-t\a_{n\!} H_0}\bigr]~.
\end{equation*}
Secondly, we retain in the expansion~(\ref{expansion}) the terms
potentially divergent at $\eps=0$. Using that $K_n(\tri,V,t)\sim
e^{-tm^2} t^{n-2}$ as $t\to0$ (for precise estimates we refer
to \cite[Section~10.2]{Polaris}), one immediately sees that
for~$n\ge3$ the corresponding contribution to the effective action is
finite, and that potential divergences only occur for~$n=1$
and~$n=2$. This is as expected, for~$n=1$ and~$n=2$ correspond to 2-
and 4-point 1PI Green functions, known to be superficially divergent,
whereas terms with~$n\ge3$ correspond to 1PI Green functions with six
or more~$\vf$, and these are known to be finite. We thus write
\begin{equation}
     \Gamma_1^{\rm reg}[\vf] = \frac{\,\mu^{2\eps}}{2} \int_0^\infty
       \frac{dt}{t^{1-\eps}}~\Tr~ \bigg[\,
         t\,V \,{\rm e}^{-tH_0} - \frac{t^2}{2} \int_0^1\! d\a ~
          V\,{\rm e}^{-t\a H_0}\, V\,{\rm e}^{-t(1-\a)H_0}
            + O(\vf^6)\,\bigg]\,.
\label{effectwofour}
\end{equation}

Let us concentrate on the contribution to the regularized effective
action quadratic in~$\vf$, given by the first term
in~(\ref{effectwofour}). Using the expressions of $H_0$ and $V$, it
takes the form
\begin{equation}
     \Gamma_1^{\rm reg\,(2)}[\vf] =  \frac{\,\la\,\mu^{2\eps}\,}{12}
       \int_0^\infty \! dt~ t^{\eps}\,{\rm e}^{-tm^2}\,\Tr\,
           \big[\,\bigl( L_{\vf\star\vf} + R_{\vf\star\vf}
            + L_{\vf}\,R_{\vf}\bigr) \,{\rm e}^{-t\tri}\,\big]\,.
\label{effecPNP}
\end{equation}
Writing the trace as
\begin{equation*}
     \Tr\,\,(\cdots) = \int d^4\!x\,\langle x\,|\,\cdots\,|\,x\rangle\,,
\end{equation*}
using Rieffel's expression (\ref{Rieffeldef}) for the star product
and recalling
\begin{equation*}
      \langle x|{\rm e}^{-t\tri}| y\rangle = \frac{1}{(4\pi t)^2}~
              {\rm e}^{-|x-y|^2/4t}\,,
\end{equation*}
it is straightforward to show that the first and second terms in the
parenthesis in~(\ref{effecPNP}) give the same contribution
\begin{equation}
     \Tr\,\big( L_{\vf\star\vf}\, {\rm e}^{-t\tri}\big) =
           \Tr\,\big( R_{\vf\star\vf}\, {\rm e}^{-t\tri}\big) =
                \frac{1}{(4\pi t)^2} \int d^4\!x \,\,\vf^2(x) \,,
\label{L-P}
\end{equation}
and the third one yields
\begin{equation}
     \Tr\,\big( L_\vf\,R_\vf \, {\rm e}^{-t\tri}\big) =
         \int\! d^4\!x \int \! \frac{d^2\!\bar{u}}{(2\pi\theta)^2}~
           \vf(\tilde{x},\bar{x})\,\,\vf(\tilde{x},\bar{x}+\bar{u}) ~
              \frac{{\rm e}^{-t\bar{u}^2/\theta^2}}{4\pi t}~.
\label{L-R-NP}
\end{equation}
It is clear that the contributions to the effective action
$\Gamma_1^{\rm reg\,(2)}[\vf]$ that result from these traces are
well-defined if $\eps$ is not an integer equal or less than one. In
fact, keeping $\eps$ away from the poles and integrating over $t$, we
obtain
\begin{equation}
\begin{array}{rl}
      \Gamma_1^{\rm reg\,(2)}[\vf]  \!\! &
         {\ds =\, \frac{\la m^2}{\,96\pi^2\,}~
          \bigg(\frac{\mu^2}{m^2}\bigg)^{\!\ds \eps}~\Gamma(-1+\eps)
              \int d^4x\,\vf^2(x) }\\[9pt]
       & {\ds +\, \frac{\la}{\,192\pi^3\,\theta^2}~
      \bigg(\frac{\mu^2}{m^2}\bigg)^{\!\ds \eps}~\Gamma(\eps)
       \int\! d^4\!x \int \! d^2\!\bar{u}
       \vf(\tilde{x},\bar{x})~ \vf(\tilde{x},\bar{x}+\bar{u}) ~\bigg(1
       + \frac{\bar{\ds u}^2}{\ds m^2\theta^2}\bigg)^{\!\!\ds-\,\eps}}\,.
\end{array}
\label{effecresult}
\end{equation}
In the first term we recognize a planar {\it local} contribution.
Hence the divergence that occurs in it as $\eps\to0$ can be
subtracted by adding a local counterterm to the classical action. This
is the usual mass renormalization. The second contribution, however,
is nonplanar and developes a {\it nonlocal singularity} as
$\eps\to0$ that can not be subtracted by a local counterterm.
This means that the theory is not renormalizable.

Let us understand this result in terms of 1PI Green funtions. By
functionally differentiating the regularized effective action we
generate the regularized 1PI 2-point Green function $\Gamma_1^{\rm
reg\,(2)}(x_1,x_2)$ as the sum
\begin{equation*}
\Gamma_1^{\rm reg\,(2)}(x_1,x_2)
       = \frac{\delta^2\Gamma_1^{\rm reg (2)}[\vf]}
              {\delta\vf(x_1)\,\delta\vf(x_2)} \bigg\vert_{\vf=0}
       = \Gamma_{\rm 1,P}^{\rm reg\,(2)}(x_1-x_2)
       + \Gamma_{\rm 1,NP}^{\rm reg\,(2)}(x_1-x_2)\,
\end{equation*}
of a planar part
\begin{equation*}
      \Gamma_{\rm 1,P}^{\rm reg\,(2)}(x_1-x_2) =
      \frac{\la m^2}{\,48\pi^2\,}~
      \bigg(\frac{\mu^2}{m^2}\bigg)^{\!\ds \eps}~\Gamma(-1+\eps)~
      \delta^{(4)}(x_1-x_2)\,
\end{equation*}
and a nonplanar part
\begin{equation*}
      \Gamma_{\rm 1,NP}^{\rm reg\,(2)}(x_1-x_2) =
      \frac{\la}{\,96\pi^3\theta^2\,}~
      \bigg(\frac{\mu^2}{m^2}\bigg)^{\!\ds \eps}~\Gamma(\eps)~
      \delta^{(2)}(\tilde{x}_1-\tilde{x}_2)~ \bigg[\, 1
       + \frac{(\bar{x}_1-\bar{x}_2)^2}{m^2\theta^2}\,\bigg]^{-\eps}\,.
\end{equation*}
The nonplanar part is neither finite nor local at $\eps=0$. This
represents a particularly nasty strain of UV divergences in the
nonplanar sector and makes the theory unrenormalizable already at one
loop. 

We next show how this result fits with the expressions in the
literature for one-loop radiative corrections in momentum space. To
this end, we need the Fourier transform of the 2-point 1PI Green
function obtained above. Restricting ourselves to the the problematic
nonplanar part, one finds when $\eps$ is not a negative
integer~\cite{Schwartz} for its Fourier transform
\begin{equation*}
     \Sigma_{\rm 1,NP}^{\rm reg}(\tilde{p},\bar{p})
          := \int d^4\!z~ {\rm e}^{-ipz}\,
            \Gamma_{\rm 1,NP}^{\rm reg\,(2)}(z)
          = \frac{\la m^2}{\,24\pi^2\,}~
            \bigg(\frac{\mu^2}{2m^2}\bigg)^{\!\!\eps}~
            \frac{\,K_{1-\epsilon}(\theta m |\bar{p}|)\,}
                 {(\theta m |\bar{p}|)^{1-\eps}}~.
\end{equation*}
Here $K_\nu(\cdot)$ denotes the third Bessel function of order $\nu$.
Apparently the limit $\eps\to0$ of this expression is harmless and one
is tempted to write
\begin{equation*}
     \Sigma_{\rm 1,NP}(\tilde{p},\bar{p}) = \frac{\,\la m^2}{24\pi^2}~
         \frac{K_1(\theta m |\bar{p}|)}{\theta m |\bar{p}|}~,
\end{equation*}
which after rotating back to Minkowski momentum space, reproduces
indeed the known results in the literature ---see e.g.~\cite{FRR}. The
key point, however, is that the function $K_1(\theta m
|\bar{p}|)/\theta m |\bar{p}|$ is not locally integrable and thus it
does not define a distribution; it has no Fourier transform either.
This matters because in the integral
\begin{equation}
      \frac{1}{2} \int \frac{d^4\!p}{(2\pi)^4}~
            \Sigma_{\rm 1,NP}(\tilde{p},\bar{p})~
                \, \hat\vf(p)\,\hat\vf(-p)\,,
\label{eff-2-momentum}
\end{equation}
which would define the corresponding contribution to the effective
action, $\hat{\vf}(p)$ is a~$c$-number function~\cite[Section~6.2]{IZ}
and such integral is undefined.

Let us discuss this point in some more detail. For a generic
$c$-number function $\hat{\vf}(p)$ the integral (\ref{eff-2-momentum})
does not exist, since
\begin{equation*}
     \frac{K_1(\theta m |\bar{p}|)}{\theta m |\bar{p}|}~  d^2\!\bar{p}
        ~\sim~ \frac{d|\bar{p}|}{|\bar{p}|}
        ~~~~{\rm for}~~~~ |\bar{p}|\to0\,.
\end{equation*}
This hints that a way to avoid the problem would be for the 2-point
1PI Green function to depend on a linear combination of $|\tilde{p}|$
and $|\bar{p}|$ so that $\Sigma_{\rm 1,NP}$ would only diverge for
both $|\tilde{p}|$ and $|\bar{p}| \to0$, in which case the modulus
$|p|$ of the four-momentum approaches zero and the measure $d^4p\sim
|p|^3 d|p|$ improves the convergence of the
integral~(\ref{eff-2-momentum}). To have such a dependence, the rank
of the noncommutativity matrix must be four instead of two. The reason
for this is that covariance implies that the 2-point 1PI Green
function in momentum space may only depend on the mass and $p\circ p
:= p_\m \Theta^{\m}_{\,\nu} \Theta^{\nu\sigma} p_\sigma$, and for
$p\circ p$ to involve $|\tilde{p}|$ and $|\bar{p}|$, a rank-four
$\Theta$ matrix is needed. Indeed, if the noncommutativity matrix has
the form
\begin{equation*}
     \Theta=\begin{pmatrix} \zeta S & 0 \\
                            0 & \theta S
            \end{pmatrix} ~,
\end{equation*}
$p\circ p$ takes the form $\zeta^2|\tilde{p}|^2 +\th^2|\bar{p}|^2$.
The 2-point 1PI Green function in configuration euclidean space can
then be calculated along the very same lines as presented here, with
the same result for the planar part, whereas for the nonplanar part
one obtains
  \begin{equation*}
      \Gamma_{\rm 1,NP}^{\rm reg\,(2)}(x_1-x_2) =
        \frac{\lambda(\mu^2\th^2\zeta^2)^\eps}{96 \pi^4}
         \frac{\Ga(1+\eps)}
            {(\th^2\zeta^2m^2 + \th^2\,|\tilde{x}_1-\tilde{x}_2|^2
		 + \zeta^2\,|\bar{x}_1-\bar{x}_2|^2)^{1+\eps}}~.
\end{equation*}
See ref.~\cite{Gayral} for the same calculation with a different
approximation method for the heat kernel and a different
regularization scheme.  Now the effective action and the Fourier
transform
\begin{equation*}
     \Sigma_{\rm 1,NP}^{\rm reg}(\tilde{p},\bar{p}) =
      \frac{\la m^2}{\,24\pi^2\,}~
            \bigg(\frac{\mu^2}{2m^2}\bigg)^{\!\!\eps}~
            \frac{\,K_{1-\epsilon}(m \sqrt{p\circ p})\,}
                 {(m \sqrt{p\circ p})^{1-\eps}}
\end{equation*}
exist for $\eps=0$. The latter depends on $|\tilde{p}|$ and
$|\bar{p}|$, thus making the integral~(\ref{eff-2-momentum})
well-defined. This formula can be Wick-rotated to Minkoswki momentum
space.

Our result holds as well for noncommutative $U(1)$ gauge theory. To
see this, recall~\cite{FRRs} on the one hand that for noncommutative
$U(1)$ gauge theory in an arbitrary Lorentz gauge the nonplanar part
of the 2-point 1PI Green function in Minkowski momentum space behaves
for arbitrary $\Theta\to 0$ as
\begin{equation}
  i\,\Pi^{\m\n}_{\rm1,NP}(\tilde{p},\bar{p})  
     = \frac{2i}{\pi^2} ~ \frac{(\Theta p)_\m (\Theta p)_\n}
                               {(p\circ p)^2}
      + \frac{i}{16\pi^2}\left(\frac{13}{3}-\alpha\right) 
        \ln\left(p^2 p\circ p\right )\,(p^2g_{\m\n} - p_\m p_\n)
      + O(\Theta^0) ,
\label{gauge}
\end{equation} 
where $\alpha$ is the gauge-fixing parameter. On the other hand, on
the grounds of covariance, we observe that $\Pi_{\m\n}$ depends on
$\bar{p}$ only through $(\Theta p)_\m$, $p^2$ and $p\circ p$. Hence,
for a space-like $\Theta$, for which $p\circ p=(\theta|\bar{p}|)^2$,
the behaviour of $\Pi_{\m\n}(p)$ as $|\bar{p}|\to 0$ is given by its
behaviour as $\theta\to 0$. It then follows from eq. (\ref{gauge})
that
\begin{equation*}
  \Pi^{\m\n}_{\rm 1,NP}(\tilde{p,}\bar{p})~ d^2\bar{p}
     ~\sim~ \frac{d|\bar{p}|}{|\bar{p}|}~
        ~~~~{\rm for}~~~~ |\bar{p}|\to 0~~~~{\rm and}~~~~\m,\n=2,3\,.
\end{equation*}
This proves once again that the integral 
\begin{equation*}
\frac{1}{2} \int \frac{d^4\!p}{(2\pi)^4}~
             A_\m(p)\, \Pi^{\m\n}_{\rm 1,NP} (\tilde{p},\bar{p})
                \,A_\n(-p)\,, 
\end{equation*} 
that would define the contribution to the effective action quadratic
in the gauge field does not exist for $A^\m(p)$ a $c$-number. The
examples exhibited in this paper suggest that this disease is quite
general for space-like noncommutative field theories with UV/IR
mixing. This result may help to explain the unability to define
quasiplanar Wick products within the framework of the Yang--Feldman
approach to quantum field theory \cite{Bahnsetal}.

In view of the state of affairs expounded here, several alternatives
to quantum field theory with space-like noncommutativity are worth
considering.
\begin{itemize}
\item
Other types of noncommutativity should be further investigated in
regard of unitarity. Some progress along this line has been made in
refs.~\cite{Bahnsetalt} and~\cite{Fujikawa}.
\item
So far we have been discussing the 2-point part of the effective
action. The 4-point part, given by the second term
in~(\ref{effectwofour}), can also be be evaluated in the same way, and
turns out to be free of this disease (although not of the UV/IR
mixing). On the face of it, one could think of performing by fiat a
non-local renormalization of the 2-point function.
\item
The pathology uncovered in this paper adds itself to an endemic list
of troubles \cite{FRRs,supersymmetry}. If one wants to stay within the
realm of conventional perturbative renormalization, it seems again
that the only way out for space-like noncommutativity is to
supersymmetrize the theory. In the case of the scalar theory, one
would expect that, in the same manner as for gauge theories the
quadratic noncommutative IR divergences in the 2-point 1PI Green
functions are cancelled by the supersymmetric partners of the gauge
field, the supersymmetric partners of the scalar field $\vf$ would
cancel the non-integrability in $\Sigma_{\rm 1,NP}^{\rm ren}$ so as to
render a well-defined contribution to the effective action.
\end{itemize}

We note finally that it has been reported~\cite{Armonietal} that, for
noncommutative gauge theory with space-like noncommutativity, the
axial anomaly acquires a non-planar contribution. A natural question
is whether the disease found here is related to this anomaly
phenomenon.

\section*{Acknowledgments}

VG wishes to acknowledge the hospitality of the Department of
Theoretical Physics of Univer\-sidad Complutense de Madrid. JMGB
thanks MEC, Spain, for support through a `Ram\'on y Cajal'
contract. He also acknowledges discussions with E.~L\'opez. FRR is
grateful to MEC, Spain, for financial support through grant
BFM2002-00950.

\end{document}